\documentclass[jcp,twocolumn,showpacs,floats]{revtex4}
\usepackage{graphicx}
\usepackage{amsmath}
\usepackage{tikz}
\usepackage{color}
\graphicspath{{images/},{images/pdf/}}
\begin{document}
\title{Adsorption of lactic acid on chiral Pt surfaces - A Density Functional Theory study}
\author{J.-H.~Franke and D.S.~Kosov} 
\affiliation{Department of Physics, Campus Plaine - CP 231, Universite Libre de Bruxelles, 1050 Brussels, Belgium}
\date{\today}

\begin{abstract}
The adsorption of the chiral molecule lactic acid on chiral Pt surfaces is studied by Density Functional Theory calculations. First we study the adsorption of L-lactic acid on the flat Pt(111) surface. Using the oPBE-vdW functional which includes van der Waals forces on an ab initio level, it is shown that the molecule has two binding sites, a carboxyl and the hydroxyl oxygen atoms. Since real chiral surfaces are (i) known to undergo thermal roughening that alters the distribution of kinks and step edges but not the overall chirality and (ii) kink sites and edge sites are usually the energetically most favored adsorption sites, we focus on two surfaces that allow qualitative sampling of the most probable adsorption sites. We hereby consider chiral surfaces exhibiting (111) facets, in particular Pt(321) and Pt(643). The binding sites are either both on kink sites - which is the case for Pt(321) or on one kink site - as on Pt(643). The binding energy of the molecule on the chiral surfaces is much higher than on the Pt(111) surface. We show that the carboxyl group interacts more strongly than the hydroxyl group with the kink sites. The results reveal rather small chiral selectivities on the order of 20 meV for the Pt(321) and Pt(643) surfaces. L-lactic acid is more stable on Pt(321)$^S$ than D-lactic acid, while the chiral selectivity is inverted on Pt(643)$^S$. The most stable adsorption configurations of L- and D-lactic acid are similar for Pt(321) but differ for Pt(643). We explore the impact of the different adsorption geometries on the work function which is important for field ion microscopy.
\end{abstract}
\maketitle

\section{Introduction}

Most biologically relevant molecules cannot be superimposed on their mirror image, i.e. they are chiral.\cite{Gellman2010} This ubiquitous feature has important consequences for biological activity of chiral molecules. Since interactions between these molecules are chirally specific, different enantiomers of drug molecules have vastly different bioactivities. To exploit this feature it is necessary to selectively synthesize one enantiomer or to separate it from a racemic mixture.\cite{Sholl2009} This, in turn requires again chirally selective interactions or chirally selective catalysts.

The standard approach to produce chiral molecules is by homogeneous catalysis,\cite{Blaser2005} which requires additional purification steps after synthesis.\cite{Sholl2009}  To avoid this additional complication it would be desirable to perform asymmetric synthesis, i.e. enantiopure synthesis directly on a surface. This necessitates a chiral surface which can be achieved by modifying it with chiral molecules.\cite{Sholl2009,Lorenzo2000,Fasel2006,Kuhnle2002,Mallat2007} Alternatively, one can hope to exploit the intrinsic chirality of high Miller index metal surfaces.\cite{Ahmadi1999,Sholl1998,Clegg2011} Since these surfaces are readily amenable to experimental surface science techniques\cite{Eralp2011,Bombis2010,Zhao2004,Greber2006} as well as Density Functional Theory (DFT) calculations\cite{Han2012,Bhatia2005,Bhatia2008,Sljivancanin2002,Greber2006} due to their well defined structure one can hope to gain a fundamental understanding of the principles underlying chiral selective properties through their study. Adsorption energetics can be studied experimentally via Temperature Programmed Desorption.\cite{Horvath2004,Huang2011,Huang2008,Cheong2011}

An interesting feature here is that chirality can also arise from achiral systems due to reduction of dimensionality. Racemic mixtures of molecules can form homochiral domains on surfaces.\cite{Bombis2010,Fasel2006,Vidal2005} Such an effect might also be implicated in the emergence of homochirality observed in biological molecules.\cite{Sowerby1998} This also highlights the qualitative differences resulting from molecule-molecule interactions in solution and collective behavior observed in molecular monolayers, an effect that might also be exploited for chirally selective catalysts.

Here we study one of the basic building blocks of green chemistry, lactic acid, on intrinsically chiral Pt surfaces.\cite{Gallezot2012,Poliakoff2002,Holm2010} Lactic acid is already produced on an industrial scale for use in food and beverages, or pharmaceuticals and can be made from renewable sources.\cite{Gallezot2012,Ragauskas2006} One especially important growth market is in its polymerized form, polylactic acid (PLA).\cite{MadhavanNampoothiri2010,Rasal2010,Katiyar2010} Here the thermochemical properties of the polymer depend also on the chirality of the monomers it was made from, which points to the importance of enantioselective control in this system.\cite{MadhavanNampoothiri2010,Platel2008} PLA of limited molecular weight can be obtained by condensation from lactic acid. To economically get to the high molecular weight polymer needed in practice this low molecular weight polymer can be broken down into lactide, the condensed dimer. This product can be purified and subsequently used as a precursor to obtain high quality high molecular weight polymer.\cite{MadhavanNampoothiri2010,Platel2008}

Specifically we focus here on the adsorption of the two enantiomers of lactic acid on Pt(321) and Pt(643) surfaces. We also study the adsorption of the molecule on the Pt(111) surface that can be considered as a model system for the terraces of the two chiral surfaces. We find that lactic acid binds to Pt surfaces predominantly through two adsorption sites: the oxygens of the hydroxyl and the carboxylic groups. On the Pt(321) surface these two binding sites of the molecule can each be bound to kink atoms, which also turns out to be the most stable adsorption configuration. On the Pt(643) surface one of them needs to bind to either a ridge atom or to be on the terrace. The most stable adsorption geometry in this case depends on the chirality of the surface and the molecule. In reality, a chiral Pt surface might undergo thermal roughening under conservation of the global chirality.\cite{Power2002,Giesen2004,Zhao2004,Baber2008} The surfaces studied here can be considered as a model system for this surface as the three surfaces offer one, two or no kink sites to bind the molecule to. 

We find a large increase in binding energy when comparing adsorption on the Pt(111) and Pt(643) surfaces and a smaller increase for Pt(643) relative to the Pt(321) surface. Analysis of the contributions of the carboxyl and hydroxyl groups to the overall binding energy on the chiral surfaces shows that the carboxylic group contributes most to the binding energy. Therefore, the additional binding energy of the hydroxyl group on the second kink site on Pt(321) is partially compensated by strain on the molecule and the carboxylic bond to the kink site. Comparing the binding energies of different molecular chiralities we find a small chiral selectivity of 23 meV and 17 meV for the Pt(321) and Pt(643) surfaces, respectively. This is comparable to other results of chiral molecules on intrinsically chiral metal surfaces.\cite{Bhatia2008,Bhatia2005,Han2012,Horvath2004,Huang2011,Huang2008} However, because L-lactic acid is more stable on Pt(321)$^S$ and less stable on Pt(643)$^S$, the overall chiral selectivity of a roughened surface is predicted to be very small. To facilitate comparison to Field Ion Microscopy (FIM) imagery we also calculate the work function changes induced upon adsorption of the molecule. This is especially interesting for the Pt(643) surface since the most stable configurations of the two enantiomers are similar in energy but have very different conformations.

The remainder of the paper is organized as follows. In Sec. 2, we describe the parameters used in the computations. Section 3 and 4 present the results obtained for adsorption on the Pt(111) and chiral surfaces, respectively. Section 5 deals with the electronic structure of the different adsorption configurations.

\section{Computational details}

\begin{figure}[Htb]
\includegraphics[width=6cm]{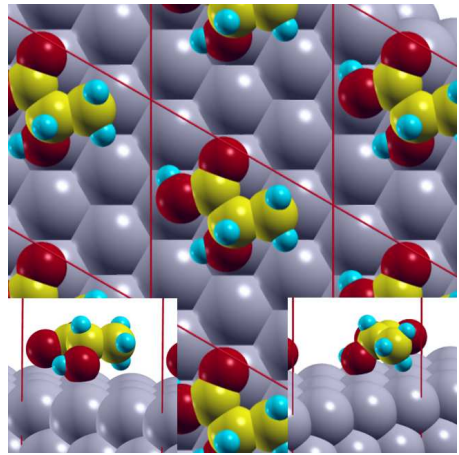}
\caption{Lactic acid adsorbed on the Pt(111) surface calculated with the oPBE-vdW functional.}
\label{fig:geom111}
\end{figure}

\begin{table*}[Htb]
\caption{Binding energies and work functions of the most stable configurations of lactic acid (L- and D-enantiomer) on the Pt surfaces studied. The first number in the binding energy column is calculated with respect to the molecule in the surface supercell (with a K-mesh like in the surface calculation). The second binding energy column gives the energy with respect to an isolated molecule in a large supercell. Referencing the binding energy to the isolated molecule reduces the calculated chiral selectivity, showing that coverage effects influence this quantity. The Hirshfeld charge of the adsorbed lactic acid molecule is given in the last column.}
\begin{tabular}{c|c|c|c|c}
\hline
\multicolumn{1}{c|}{lactic acid on} & \multicolumn{2}{|c|}{binding energy oPBE-vdW with respect to} & \multicolumn{1}{c}{work function} & \multicolumn{1}{|c}{Hirshfeld charge} \\
 & molecule in surface unit cell (eV) & isolated molecules (eV) & (eV) & of molecule (e)\\
\hline
Pt(111)            & -0.803 & -0.838 &  4.67 & 0.14 \\
L on Pt(321)$^S$   & -1.297 & -1.288 &  4.59 & 0.25 \\
D on Pt(321)$^S$   & -1.274 & -1.270 &  4.61 & 0.26 \\
L on Pt(643)$^S$   & -1.232 & -1.283 &  4.92 & 0.17 \\
D on Pt(643)$^S$   & -1.249 & -1.289 &  4.57 & 0.17 \\
\hline
\end{tabular}
\label{tab:energ}
\end{table*}

\begin{table*}[Htb]
\caption{Binding energy component analysis for the chiral surface configurations. Binding energies of hydrogen saturated COH and COOH groups in the frozen adsorption geometries and deformation energies of the substrate and molecule are calculated. It is evident that the deformation energy is larger for the more strongly interacting Pt(321)$^S$ surface when compared to Pt(643)$^S$. The COOH group is bound more strongly than the COH group throughout, an effect that is much more pronounced on the Pt(643)$^S$ surfaces where the COH group is either far away from the surface (L-lactic acid) or bound to the (111) facet (D-lactic acid). For Pt(321)$^S$ the COH group is bound to a kink atom and it is much more strongly interacting than on Pt(643). The sum of the binding energy components considered is smaller than the actual binding energy (cf. Table \ref{tab:energ}) which might stem from the binding energy of the neglected carbon and hydrogen atoms.}
\begin{tabular}{c|c|c|c|c|c}
\hline
\multicolumn{1}{c}{lactic acid on} & \multicolumn{2}{|c}{deformation energy (eV) } & \multicolumn{2}{|c}{binding energy (eV)} & \multicolumn{1}{|c}{sum of components (eV)} \\
 & surface & molecule & COH group & COOH group &    \\
\hline
L on Pt(321)$^S$    & 0.119 & 0.202 & -0.632 & -0.911 & -1.232  \\
D on Pt(321)$^S$    & 0.084 & 0.150 & -0.717 & -0.733 & -1.219  \\
L on Pt(643)$^S$    & 0.073 & 0.094 & -0.074 & -1.154 & -1.112  \\
D on Pt(643)$^S$    & 0.077 & 0.086 & -0.219 & -1.098 & -1.194  \\
\hline
\end{tabular}
\label{tab:decomp}
\end{table*}

We obtained our results with the DFT code VASP 5.12\cite{Hafner2008,Kresse1996a,Kresse1996b} with the oPBE-vdW functional\cite{Klimes2010,Klimes2011} throughout. The inclusion of the van der Waals forces is crucial for the adsorption of a weakly bound molecule since they dominate the binding energy, which was found by comparing to calculations using the PBE functional\cite{Perdew1996}. We opted for this special version of the vdW-DF functional since we wanted to keep the PBE class of functionals while also aiming at an optimal accuracy with the vdW nonlocal correlation. The Projector Augmented Wave method\cite{Bloechl1994,Kresse1999} was employed with valence wave functions expanded up to an energy cutoff of 400 eV. The lattice constant of Pt was determined using a K-mesh of 17x17x17 in conjunction with the tetrahedron method with Bloechl corrections. Fitting of a series of fixed volume calculations to the Murnaghan equation of state gave a lattice constant of 3.999\AA\ (3.978\AA\ for the PBE functional). All structural relaxations are carried out until all forces are smaller than 10 meV/\AA. For all slab calculations dipole corrections to the potential are applied throughout.\cite{Neugebauer1992} To increase accuracy all energies given are calculated with evaluation of the projector functions in reciprocal space.

The Pt(111) is constructed as a 6-layer slab with the two topmost layers relaxed. The K-mesh was sampled with a 17x17x1 K-mesh and Gaussian broadening of the energy levels of 0.1 eV to facilitate convergence. Pt(321) and Pt(643) surfaces were constructed with a thickness corresponding to 6 layers of Pt(111) and the upper half of the slabs were relaxed using 7x7x1 and 5x5x1 K-meshes, respectively. The general relaxation pattern is one of inward relaxing step edges and upward moving Pt atoms directly under the step edges for the chiral surfaces. It is similar for both PBE and oPBE-vdW functionals. 

The molecules were adsorbed on a 3x3 supercell of Pt(111) and a 2x2 supercell of Pt(321), while for Pt(643) a single surface unit cell was used. Due to the increased unit cell size the K-mesh was reduced to 3x3x1 for (3x3)-Pt(111) and (2x2)-Pt(321), respectively. All molecular degrees of freedom were allowed to relax as was the upper part of the metal slab. In the case of the chiral surfaces the molecules were adsorbed on the relaxed side of the surfaces which constitutes a Pt(321)$^S$ and a Pt(643)$^S$ surface.\cite{Ahmadi1999,Sholl2001} Adsorption energies $E_{adsorption}$ are given with reference to the isolated surface $E_{surface}$ relaxed upon removing the molecule from the unit cell using identical computational parameters and the energy of the molecule $E_{mol}$

\begin{equation}
E_{adsorption} = E_{mol\ on\ surface} - E_{surface} - E_{mol}.
\end{equation}
Two different values for $E_{mol}$ are used: (i) the energy of the molecule relaxed in the surface unit cell and (ii) the energy of the isolated molecule in a larger unit cell. The binding energy with respect to (i) thus removes molecule-molecule interactions from the adsorption energy while the calculation with respect to (ii) does not.

\section{Lactic acid on Pt(111)}

To gain insight into the adsorption behavior of lactic acid on Pt surfaces we first adsorbed the molecule on a Pt(111) surface. This surface can also be considered as a model for the (111) terraces exhibited by the Pt(321) and Pt(643) surfaces. The relaxation of the L-Lactic acid on Pt(111) yielded an adsorption energy of 0.838 eV. Comparing to the adsorption energy of 0.161 eV predicted by a separate calculation using the PBE functional this points to the importance of van der Waals forces in this system. The distance between the hydroxyl and carboxylic oxygen atoms and the nearest Pt atoms are 2.50\AA\ and 3.16\AA, respectively. These distances are predicted to be very similar (2.55\AA\ and 3.23\AA) by the PBE functional. However, the carboxylic oxygen atoms are in a plane parallel to the surface in the case of the oPBE-vdW functional (see Fig. \ref{fig:geom111}), while for the PBE functional one oxygen is further away from the surface. Nevertheless this shows that despite the much larger binding energy the geometry is similar for the two functionals.

\section{L-lactic acid on Pt(321) and Pt(643)}

\begin{figure*}[Htb]
\includegraphics[width=12cm]{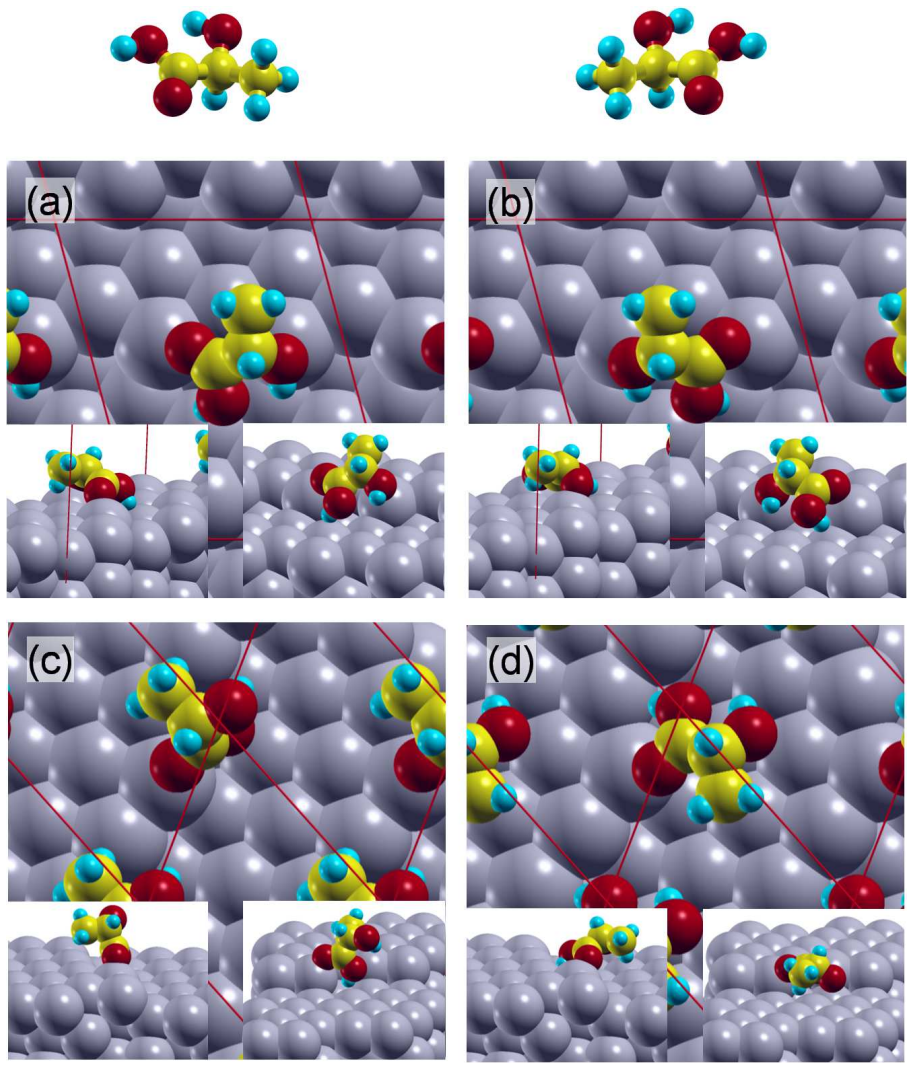}
\caption{Most stable adsorption configurations of L-lactic acid (\textbf{a},\textbf{c}) and D-Lactic acid (\textbf{b},\textbf{d}) on the chiral Pt(321)$^S$ (\textbf{a},\textbf{b}) and Pt(643)$^S$ (\textbf{c},\textbf{d}) surfaces calculated with the oPBE-vdW functional.}
\label{fig:geom}
\end{figure*}

\begin{figure*}[Htbp!]
\includegraphics[width=14cm]{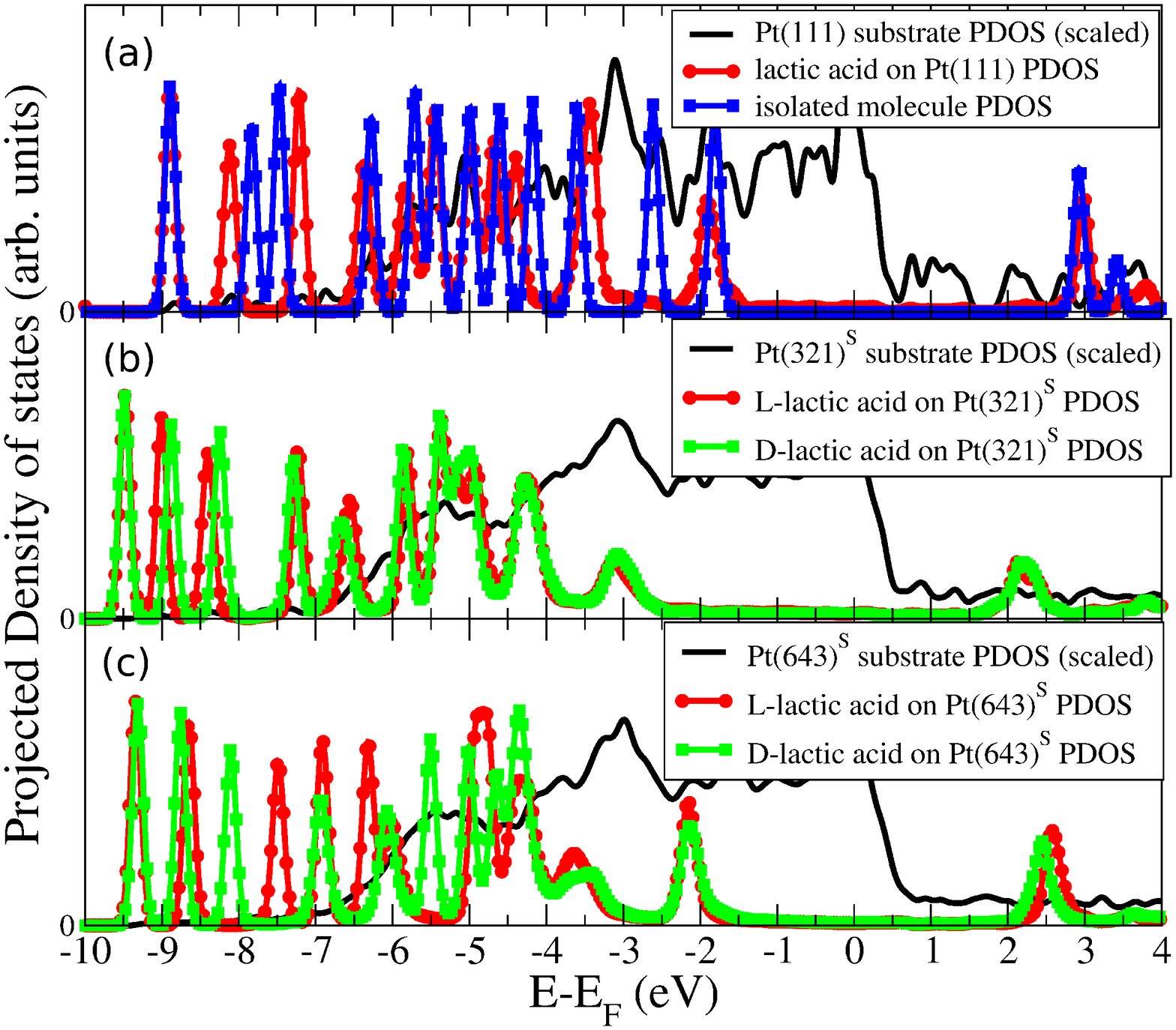}
\caption{All orbitals are projected onto atomic spheres of all atoms in the unit cell giving the Projected Density of states (PDOS). Summing over all PDOS values for all atoms belonging either to the lactic acid molecule or the substrate then gives the PDOS of the molecule and substrate, respectively. The figure shows the PDOS of the adsorption configurations of lactic acid on (\textbf{a}) Pt(111), (\textbf{b}) Pt(321)$^S$ and (\textbf{c}) Pt(643)$^S$. For the Pt(111) surface the PDOS of an isolated molecule configuration is given for comparison, while for (\textbf{b}) and (\textbf{c}) the PDOS for the two enantiomers of lactic acid are shown. It is evident that the HOMO orbital is broadened and shifted to higher binding energies for adsorption on Pt(643) and especially on Pt(321) with respect to adsorption on Pt(111). The gap between HOMO and HOMO-1 is also widened by a similar amount for all molecule-surface configurations when compared to the molecule in vacuum in (\textbf{a}). However, the HOMO-1 peak broadening is stronger for the Pt(643) surface than for the Pt(321) surface. Also evident is the close resemblance of the PDOS of different enantiomers on a given surface. The impact of chirality matching on the electronic structure of the molecule is thus limited, even for the case of different adsorption configurations on Pt(643)$^S$.}
\label{fig:PDOS}
\end{figure*}

On the chiral surfaces Pt(321)$^S$ and Pt(643)$^S$ we studied the adsorption of both enantiomers of lactic acid to gain insight into differences between their adsorption behavior. Initial positional sampling showed that the lactic acid molecule adsorbs preferentially with its oxygen binding sites on the kink sites of the surfaces. Calculated binding energies are much higher for binding to kink atoms, which is also consistent with findings from Temperature Programmed Desorption experiments for (R)-Methylcyclohexanone on chiral Cu surfaces.\cite{Huang2008} Therefore, we tested for each surface-enantiomer combination a set of configurations with either the hydroxyl or carboxyl group above the kink site. Rotating the molecule around this binding site in steps of 60 degrees then yielded the starting configurations from which the structural relaxations were carried out. Thus, for each chirality of the molecule and each chiral surface 12 configurations are considered. On the Pt(321) surfaces, however, one adsorption configuration has both molecular binding sites at kink sites. Since this constitutes at the same time the 0 degree configuration of the hydroxyl and carboxyl on kink series the overall number of configurations studied on Pt(321) for each chirality is reduced to 11. Thus, we calculated 24 configurations for Pt(643) and 22 for Pt(321) for an overall 46 structural relaxations.

The binding energies calculated are significantly higher on the chiral surfaces then on the flat Pt(111) surface (cf. Table \ref{tab:energ}). The Pt(321)$^S$ surface allows for simultaneous binding to two kink sites for both hydroxyl and carboxyl oxygen atoms (see fig. \ref{fig:geom}) which turns out to be the most favorable adsorption site for both chiralities.

In the case of Pt(643)$^S$ the most stable adsorption configurations are with the carboxyl oxygen atom bound to the kink sites. For L-lactic acid the most stable configuration has the molecule standing upright above the kink site with the carboxyl-hydroxyl carbon bond almost parallel to the surface normal so that it exhibits no hydroxyl oxygen bond to the surface. For D-lactic acid the most stable configuration is lying almost flat on the surface. Still, the hydroxyl oxygen is at a distance of 3.7 \AA\ to the nearest Pt atom, so this interaction also seems to be weak. Thus, for lactic acid on the Pt(643)$^S$ surface the hydroxyl-oxygen-surface interaction does not seem to play a role in the most stable configurations, while the carboxyl group is bound to the kink site.

The overall chiral selectivity of the surfaces studied is very small. The energy differences between the configurations of the two different chiralities are only 23 and 17 meV for Pt(321)$^S$ and Pt(643)$^S$, respectively, when referencing the energy to the molecules in the surface unit cell. Referencing instead to the energy of the isolated molecule gives reduced chiral selectivities which points to a coverage dependency of this quantity. This reduction is more pronounced in the case of Pt(643)$^S$. On Pt(321)$^S$ L-lactic acid is more stable, while on Pt(643)$^S$ the D enantiomer is the more stable one. The bond lengths of the different oxygens to the Pt atoms range from 2.29 \AA\ to 2.37 \AA\ in the case of Pt(321)$^S$. On the Pt(643)$^S$ surface similar bond lengths on the kink site are observed - 2.16 \AA\ and 2.18 \AA\ for the carboxylic oxygen. However, the bond lengths to the terrace Pt atoms are significantly longer in this case.

To understand the role of the carboxyl and hydroxyl groups in the adsorption process we carried out an energy decomposition analysis along the lines of Ref. \onlinecite{Sljivancanin2002} (see Table \ref{tab:decomp}). For each adsorption configuration the molecule is removed apart from the functional group whose binding energy contribution is to be evaluated. All atoms are held fixed in these calculations apart from the hydrogen atom introduced to saturate the bond of the functional group to the rest of the removed molecule. Also the deformation energy of the substrate and molecule are calculated by evaluating their energy at their frozen adsorption geometry upon removal of the other part, i.e. the molecule or the substrate, respectively. Since the binding energies of the molecular fragments are calculated in their frozen geometry, the sum of the binding energy of all components and the relaxation energies of the molecule and the substrate should give the adsorption energy of the whole molecule. The difference obtained in practice can be taken as a measure of the quality of the approximation made in decomposing the energies in this way. 

First of all, the relaxation energies obtained are larger for the Pt(321)$^S$ configurations than for the Pt(643)$^S$ ones. This is in line with the more strongly interacting molecules being bound to two kink sites (cf. Table \ref{tab:decomp}). Secondly, the adsorption energy of the carboxyl group is larger than the one of the hydroxyl group for all configurations. On Pt(321)$^S$ the kink-bound hydroxyl binding energy is also sizable while it is much smaller on Pt(643)$^S$. However, the carboxyl group is generally the dominant binding site. In the case of L-lactic acid on Pt(643)$^S$ the energy contribution of the hydroxyl group is especially small which can be attributed to the large distance from the surface. For D-lactic acid on Pt(643)$^S$ it is interacting with the facet which yields an adsorption energy contribution larger than for the L-lactic acid configuration but much smaller than on the kink-bound hydroxyl oxygens on Pt(321)$^S$. Interestingly, the carboxyl group adsorption energy is largest for the Pt(643)$^S$ configurations as this bond can be optimized due to the much smaller specificity of the hydroxyl bond when compared to the Pt(321)$^S$ configurations. The large difference between the binding energy of the carboxyl group on the kink and the hydroxyl group on the facet explains how it is possible to detach this group from the surface for the most stable configuration of L-lactic acid on Pt(643)$^S$.

\begin{figure}[Htb]
\includegraphics[width=9cm]{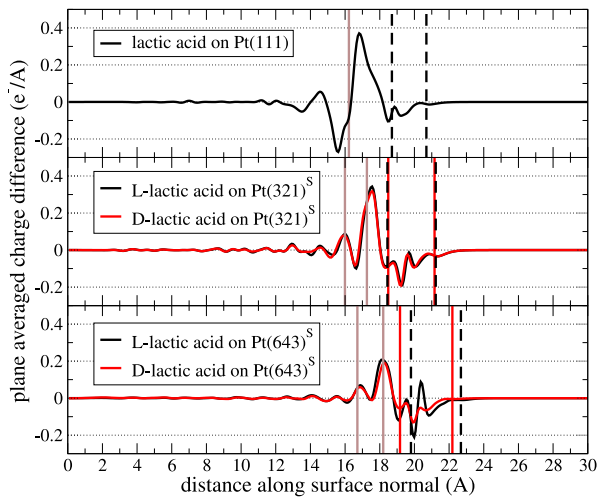}
\caption{Plane-averaged charge density redistributions upon adsorption of the lactic acid molecule enantiomers on the different Pt surfaces. The substrate is at low z-values and a vertical brown line marks the height of the uppermost surface atom in the case of Pt(111) and the highest and lowest z-values of the first layer atoms in the case of the stepped Pt(321) and Pt(643) surfaces. The highest and lowest atoms of the molecules are also marked by black dashed lines for L-lactic acid and and red lines for D-lactic acid. The observed charge redistribution pattern is very similar for all configurations which attests to the domination of the push-back effect in these configurations.}
\end{figure}

\section{Electronic structure}

To understand the bonding patterns of the lactic acid molecule on the different Pt surfaces we analyzed the electronic structure of the molecule on the different surfaces. To this end, we studied the Projected Density of States (PDOS) for the different surface-enantiomer combinations.\cite{Hoffmann1988} We also studied the charge density redistribution pattern on the surface and calculated the work function for all configurations.

Fig. \ref{fig:PDOS} shows that due to the adsorption process there are some distinct changes in the PDOS of the lactic acid molecule. In general the peaks corresponding to the highest occupied and lowest unoccupied molecular orbitals (HOMO and LUMO) are broadened with the smallest broadening occurring for the molecule adsorbed on the Pt(111) surface and the largest occurring for the Pt(321) adsorbed configurations. The HOMO is also shifted to higher binding energies with the largest shift occurring on the Pt(321) surface and the smallest one on the Pt(643) surface. Also, the gap between HOMO and HOMO-1 is widened with the HOMO-1 being shifted to higher binding energies for all adsorbed molecules. These changes in the PDOS indicate that the bonding process involves a rehybridization of the HOMO and HOMO-1 states with electronic states of the surface. Also, a look at the spatial distribution of the frontier orbitals of the molecule shows electronic density on the binding sites, i.e. the hydroxyl and carboxyl groups. This is consistent with local interaction at two binding sites on the Pt(321) surface, one binding site on Pt(643) and generally smaller interaction on Pt(111) leading to progressively smaller broadening of the frontier orbital peaks.

One particularly important experimental observable is the work function of the different surface-adsorbate configurations.\cite{Ishii1999} Upon adsorption of the molecule the distribution of the electron density of the surface is altered by the presence of the molecule. These changes are made up of a push-back effect of electron density towards the surface by the Pauli repulsion exerted on the surface electron density by the atoms of the molecule and the local charge redistribution due to the formation of chemical bonds.\cite{Bagus2002,Michaelides2003} To get an overview of the charge redistribution pattern, we calculate the charge density change $n_{diff}(r)$ upon molecular adsorption as

\begin{equation}
n_{diff}(r)=n_{ads. mol.}(r)-n_{mol}(r)-n_{subst}(r).
\end{equation}
with $n_{subst}(r)$, $n_{mol}(r)$ and $n_{ads. mol.}(r)$ denoting the electron density of the complete system, molecule and of the substrate, respectively. The charge densities of the molecule and the substrate are calculated at the frozen geometry of the molecule-surface system with either the molecule or the surface removed. It is then averaged in planes perpendicular to the surface normal to get the net contribution of the charge density rearrangement to the surface dipole which generates the work function changes.

It is found that the charge redistribution pattern for all adsorption configurations exhibits a common feature. There is significant charge accumulation just above the surface and significant charge depletion on the molecule. This can be attributed to the push-back effect. The electron density above the pristine surface is pushed back towards the surface because of the Pauli repulsion of the electrons due to the presence of the molecule. The fact that the electron redistribution is very similar for all adsorption configurations attests to the minor importance of chemical bonding in the charge redistribution patterns. While the Pt(321) configurations give quasi-identical plane averaged charge redistribution patterns, there are differences for the Pt(643) configurations. For L-lactic acid another dipole layer of opposite sign is modulated on top of the general charge redistribution pattern. This increases the work function for this particular adsorption configuration by 0.3 eV. We attribute its appearance to a reduced Push-back effect as a result of the more upright adsorption configuration.

\section{Conclusion}

We studied the adsorption of the chiral molecule lactic acid on the Pt(111), Pt(321) and Pt(643) surfaces. We found that the molecule adsorbs most strongly on the surface exhibiting the highest density of kink sites, which is the Pt(321) surface, closely followed by the Pt(643) surface. On the closely packed Pt(111) the adsorption energy is significantly lower. The lactic acid molecule hereby shows a tendency to bind with its carboxyl group to the kink sites of the chiral surfaces. On Pt(321), also the hydroxyl group is adsorbed on a neighboring kink site, while the adsorption geometry in the case of Pt(643) depends on chirality. For the Pt(643)$^S$ surface and D-lactic acid the molecule lies on the (111) facet of the surface, while for L-lactic acid on Pt(643)$^S$ the molecule stands upright on the kink site.

The chiral selectivity calculated is small at about 20 meV for the Pt(321) and Pt(643) surfaces, when referencing the energy to the molecules in the surface unit cell (without substrate) but even smaller when reference to the energy of an isolated molecule is made. However, the calculated chiral selectivity has opposite sign for the Pt(321) and Pt(643) surfaces, i.e. L-lactic acid is more stable on Pt(321)$^S$ and less stable on Pt(643)$^S$. Experimental observation of an overall chiral selectivity on a real chiral Pt surface vicinal to the (111) surface is thus predicted to be challenging, though possible.\cite{Huang2011,Huang2008} Analysis of the contributions of the carboxyl and hydroxyl groups to the total binding energy shows that the carboxyl group is the dominant binding site, giving the biggest binding energy contributions.

The adsorption process leads to a rehybridization of the frontier orbitals with electronic states of the surface. This effect is more pronounced for the most strongly bound configurations on Pt(321), less so for Pt(643) and least pronounced for Pt(111). The charge redistribution of the surface due to the adsorption of the lactic acid molecule shows the hallmark of the push-back effect that pushes electron density closer to the surface due to the Pauli repulsion of the molecular electrons. This leads to a considerable lowering of the work function to values around 4.6 eV for all molecule-surface combinations with the lactic acid on Pt(111) surface showing a work function of 4.7 eV. An outlier in terms of work function is the L-lactic acid on Pt(643)$^S$ combination. Here the upright molecular configuration leads to a smaller push-back effect that in turn yields a higher work function of about 4.9 eV.

Overall our results show that lactic acid adsorbs on stepped Pt surfaces predominantly through bonding of its carboxyl group to a kink site with the hydroxyl group constituting a secondary binding site. A small chiral selectivity on chiral Pt surfaces is predicted, whose sign depends on the exact surface studied. Adsorption geometries can depend on molecular chirality, leading to large changes in the work function. Especially this last effect should be verifiable by Field Ion microscopy or Scanning Tunneling Microscopy.

\section{Acknowledgements}

This work has been supported by the Francqui Foundation, and Programme d'Actions de Recherche Concertee de la Communaute Francaise, Belgium. We would like to thank Pierre Gaspard and Thierry Visart de Bocarme for useful discussions. We also acknowledge the Computing Center of ULB/VUB for computer time on the HYDRA cluster.


\end{document}